\documentclass[12pt]{iopart}
\usepackage{iopams}
\usepackage{setstack}
\usepackage{verbatim}
\usepackage{graphicx}
\usepackage{xcolor}
\usepackage{soul}
\usepackage[normalem]{ulem}

\newcommand{\pol}{\hat{\bf e}}

 \newcommand{\bra}[1]{\langle
#1  | } \newcommand{\ket}[1]{ | #1 \rangle  }

\newcommand{\V}[1]{{\bf #1}}

\newcommand{\expct}[1]{ \langle #1\rangle}

\newcommand{\eqref}[1]{(\ref{#1})}
\long\def\symbolfootnote[#1]#2{\begingroup\def\thefootnote{\fnsymbol{footnote}}
\footnote[#1]{#2}\endgroup}

\begin{document}

\title[Bragg spectroscopic interferometer and measurement-induced correlations in BECs]{Bragg spectroscopic interferometer and quantum measurement-induced
correlations in atomic Bose-Einstein condensates}

\author{M D Lee$^{1}$\footnote[1]{These authors contributed equally to
this work.}, S Rist$^{2}\dagger$ and  J Ruostekoski$^1$}

\address{$^1$ School of Mathematics, University of Southampton,
Southampton, SO17 1BJ, UK}
\address{$^2$ CNR-SPIN, Corso Perrone 24, I-16152 Genova and NEST, Scuola Normale Superiore, I-56126 Pisa, Italy}
\ead{mark.lee@soton.ac.uk}

\begin{abstract}
We theoretically analyze the Bragg spectroscopic interferometer of two spatially
separated atomic Bose-Einstein condensates that was experimentally realized by
Saba {\it et al.} [{\it Science} 2005 {\bf 307} 1945] by continuously monitoring
the relative phase evolution. Even though the atoms in  the light-stimulated
Bragg  scattering interact with intense coherent laser beams, we show that the
phase is created by quantum measurement-induced back-action on the homodyne
photo-current of the lasers, opening possibilities for quantum-enhanced
interferometric schemes. We identify two regimes of phase evolution: a running
phase regime which was observed in the experiment of Saba {\it et al.}, that is
sensitive to an energy offset and suitable for an interferometer, and a trapped
phase regime, that can be insensitive to applied forces and detrimental to
interferometric applications.
\end{abstract}


\date{\today}

\section{Introduction}

Bragg spectroscopy has become an established spectroscopic tool in ultracold
atom experiments
\cite{Ketterle_Bragg,Phillips,Davidson,Zimmermann_Bragg,Inguscio,Sengstock}. In typical
set-ups an intersecting pair of low-intensity pulsed laser beams is used to
excite atoms to higher momentum states. The momentum kick experienced by the
atoms corresponds to the recoil of a photon upon light-stimulated scattering
between the two laser beams. As the spontaneous scattering for off-resonant
lasers is negligible and the photons are only exchanged between the directed
coherent laser beams, the momentum transfer of the atoms can be measured for
specific values of the energy and the momentum. In particular, in a
spectroscopic analysis of the many-particle properties of ultracold atoms it is
sufficient in the scattering process to describe the light beams classically.

In the experiments by Saba {\it et al.} \cite{Saba} the relative phase coherence between two
Bose-Einstein condensates (BECs) was measured by Bragg scattering atoms between two condensate
fragments. Previous Bragg spectroscopy experiments based on time of flight had concentrated on
directly detecting the atoms that were transferred to higher momentum states by the laser beams.
In the experiment by Saba {\it et al.} \cite{Saba}, however, the strength of the Bragg
scattering was measured by monitoring the variations of the light intensity in the laser beams
by homodyne detection. Due to the correspondence between the light-stimulated scattering of
photons between the laser beams and the atoms scattered between two momentum states, the
intensity fluctuations are directly proportional to the number of atoms scattered between the
condensate fragments.
\begin{figure}[htp] \centering
\includegraphics[width=0.5\textwidth]{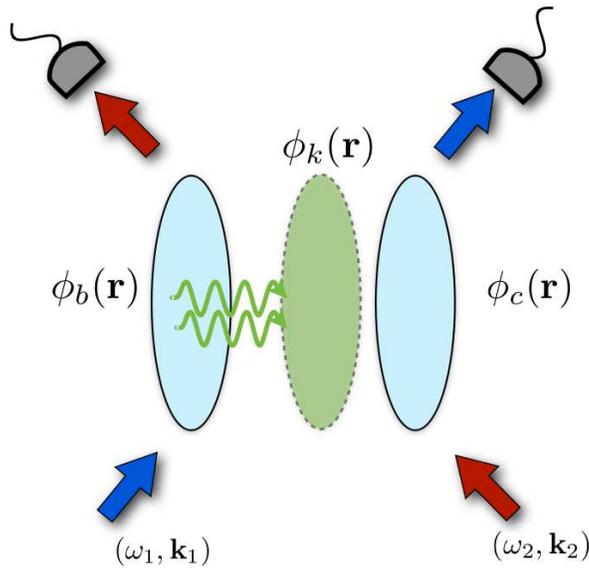}
\caption{ \label{fig:Setup} Our model of the Bragg interferometric
measurement of the relative phase
between two distant BECs.  The two BECs are described by the
macroscopic wavefunctions $\phi_b(\V{r})$ and $\phi_c(\V{r})$, and are illuminated by two
coherent laser beams.  Bragg scattering imparts momentum to atoms from the left condensate,
transferring them to the state described by
$\phi_k(\V{r})$.  After an appropriate time the outcoupled atoms will overlap with the right
condensate, and the Bragg beams will drive Rabi oscillations between the two atomic
clouds. This establishes an optical weak link between the two BECs, and
continuous monitoring of the intensity fluctuations in the laser beams measures
the phase coherence between the BECS.}
\end{figure}

Saba {\it et al.} \cite{Saba} measured the light intensity variations of the
Bragg beams which revealed relative phase coherence between the condensates
even when the BECs were independently produced and possessed no a priori phase
information. By analyzing theoretically a continuous atom detection process it
has been previously shown that the back-action of quantum measurement of the
atomic correlations
\cite{Javanainen_BEC,Cirac_ContObs,JAC96,WON96,Yoo,Castin_RelPhase,Dunningham1999,Laloe2007},
and analogous photon correlations \cite{Janne_NonDest,MacSupJanne}, can
establish a relative phase between two BECs even when they  have `never seen
each other' before. It has also been suggested that phase-coherent states of
condensates may naturally emerge as robust state descriptions due to
dissipative interaction with the environment \cite{BAR96,ZUR91}.  With regards
to the Bragg spectroscopic interferometer of \cite{Saba}, the question we
ask is: {\it how is the phase coherence between the two BECs created, given
the condensates interact with coherent laser beams that can usually be
described classically?}

Here we analyze a model of the experimental detection scheme~\cite{Saba},
illustrated in figure~\ref{fig:Setup}, and show that the phase coherence can be
built up by continuously monitoring the photo-current obtained by a homodyne
measurement that describes the intensity fluctuations of the laser beams.  We
show a rapid establishment of a well-defined relative phase between two
independently produced BECs. We identify two distinct regimes of  subsequent
phase evolution: a running phase and a trapped phase regime. In the running
phase regime the relative phase grows linearly proportional to the energy offset
between the two condensate wells and could be suitable for a weak force
detection in interferometric applications \cite{Saba}. In  the trapped phase
regime, in the case of a very weak energy offset, the measurement process drives
the system close to a dark state where a destructive interference between
different scattering paths suppresses the intensity fluctuations of the lasers.
In the trapped phase regime the effect of the energy offset on the phase
evolution is suppressed, potentially to the detriment of interferometric
applications.

Our analysis demonstrates how Bragg spectroscopy can be sensitive to subtle
quantum features of ultracold atom systems. Quantum measurement-induced
back-action of the photo-current detection on the relative phase coherence of
the BECs represents a spatially non-local entanglement of the laser beams and
the relative many-particle state of the atoms. Indeed, the location of the
photo-current detection can be far separated from the interaction region of the
coherent laser beams and the atoms. Moreover, one Bragg pulse can be used to
entangle the two spatially isolated BECs. A second pulse may then be employed in
optical readout of the subsequent evolution dynamics of the
measurement-established relative phase coherence between the condensates. An
energy offset between the two condensate wells between the subsequent pulses
would result in a detectable phase shift providing potential interferometric
applications \cite{Saba}. Here the phase is determined by a continuous quantum
measurement process opening possibilities for quantum feedback and control
methods, e.g., in generation of sub-shot-noise phase-squeezed
states~\cite{Wiseman_book,Feedback}. Such states may be useful in
quantum-enhanced  metrology in the realization of a high-precision quantum
interferometer overcoming the standard quantum  limit of classical
interferometers \cite{wineland, bouyer, holland,giovannetti}.   Probe field
response was also recently measured in Bragg spectroscopy of condensate
excitations in a heterodyne-based detection system, which was able to reach the
shot-noise limit~\cite{Cornell2011}. Previous theoretical studies of the
effects of continuous monitoring on  light scattered from BECs have considered
photon counting \cite{Janne_NonDest,MacSupJanne}, e.g., in preparation of
macroscopic superposition states \cite{MacSupJanne}, and dispersive
phase-contrast imaging \cite{dalvit,hope}, e.g., in suppression of heating
\cite{hope}.

The article is organized as follows.
In section~\ref{sec:Model} we give a short review of the experimental
setup of \cite{Saba} and the relevant results for this work. We then introduce our
basic theoretical  model.
In section~\ref{contmeas} we derive a stochastic differential equation which describes
the evolution of the system under the continuous measurement of the scattered light
intensity. In section~\ref{numres} we present our numerical results with a physical
interpretation. Finally, some concluding remarks are made in section~\ref{Sum}.

\section{Model and effective Hamiltonian} \label{sec:Model}

An interferometric scheme between two spatially isolated BECs was experimentally
realized in \cite{Saba} without the need for splitting or recombining the
two condensate atom clouds. The method was based on stimulated light scattering
of a small fraction of the atoms, only weakly perturbing the condensates and
therefore representing an almost nondestructive measurement. Two isolated BECs
were prepared in the sites of an unbalanced double-well potential, and
illuminated by the same pair  of Bragg beams.  These beams outcoupled atoms from
each well, and interference between such atoms provided a coupling between the
BECs. When outcoupled atoms from one condensate spatially overlapped the second,
measurement of the Bragg beam intensity was shown to be sensitive to the
relative phase $\Phi$ between the condensates.  In addition, the potential
offset between the two wells gave rise to a difference in energies
$\delta \!\mu$, which in turn led to a relative phase evolution $\Phi(t) =
\Phi(0)+\delta\!\mu \,  t/\hbar$. This was observed as oscillations in
the Bragg beam intensity of frequency $\omega_{\rm osc}=\delta\! \mu/\hbar$,
demonstrating how monitoring the Bragg beam intensity directly measured the
dynamical evolution of the relative phase between the macroscopic wavefunctions.

In the experiment a single Bragg pulse established a random relative phase
between the two independently produced BECs. If two successive Bragg pulses were
applied to the same BEC pair the relative phase measured by the second pulse
was correlated with that detected by the first pulse, indicating
that the interaction of the first Bragg beam with the atoms had projected the
system into a state with a well-defined relative phase between the condensates.

The key to the method is the weak link established between the BECs by the Bragg
laser beams that couple out small atomic samples from the condensates
\cite{Shin_OptWeakLinkBEC}. The coherently driven population dynamics between the BECs
is influenced by the relative phase coherence~\cite{JAV96b,ruo_phasediff}, and the Bragg
scattering may be understood as an interference in momentum space~\cite{Pitaevskii99}.
The specific advantage of the Bragg spectroscopic
interference scheme \cite{Saba} is the nondestructive nature of the detection
process, potentially constituting a major advance for interferometric applications
since it allows one to probe the evolution of the phase coherence in time by a
continuous measurement process
\cite{Javanainen_comment,AtomInt_revA,AtomInt_revB}. It has also
been argued that this setup can be seen as an analog to homodyne detection for
matter waves \cite{Rist_homodyne}.

In order to analyze the continuous measurement process of Bragg spectroscopy we
consider the system depicted in figure~\ref{fig:Setup} that is analogous to the
experimental set-up of \cite{Saba}. We assume that the two condensates are
initially uncorrelated and that there is no tunnelling between the two spatial
regions. As in the experiment, an offset in the trapping potential between the two
condensates is accounted for by a difference in chemical potential, $\delta\!\mu$.
The condensates are illuminated by two Bragg beams, which impart momentum, kicking
atoms out of the traps.  The outcoupled atoms propagate from the left condensate to
the right and establish an optical weak link between the two macroscopic
wavefunctions \cite{Shin_OptWeakLinkBEC}.

For simplicity, in the theoretical analysis we use a single mode approximation for
the condensates and assume that all atoms in the left (right) condensate are in
the state $\ket{b}$ ($\ket{c}$). The atoms in the left (right) condensate are then
described by the second quantized field operators
$\hat{\psi}_L(\V{r})=\phi_b(\V{r}) \hat{b}$ ($\hat{\psi}_R(\V{r})=\phi_c(\V{r})
\hat{c}$) which fulfill the usual bosonic commutation relations. Here $\hat{b}$
($\hat{c}$) annihilates an atom in state $\ket{b}$ ($\ket{c}$) and
$\phi_{b,c}(\V{r})$ obey the Gross-Pitaevskii equation \cite{Stringari_Book}.

Atoms from the BEC in the left well in state $\ket{b}$ are transferred by the Bragg
beams to the momentum state $\V{k}=\V{k}_1-\V{k}_2$, where $\V{k}_j$ are the
wavevectors of the Bragg beams. The outcoupled atoms propagate with momentum $\V{k}$
towards the right BEC in state $\ket{c}$. We take the
wavefunction of the outcoupled atoms $\phi_k(\V{r})$ to be the momentum shifted
original wavefunction
\begin{equation}
\phi_k(\V{r}) = \phi_b(\V{r}-\V{r}_L)
\rme^{\rmi \V{k}\cdot (\V{r}-\V{r}_L)} \, ,
\end{equation}
where $\V{r}_L$ $(\V{r}_R)$ gives the position of the centre of the left (right)
trap.  We assume that the momentum kick of the atoms is sufficiently strong, so
that the essential characteristics of the continuous quantum measurement process
are not obscured by collisions with the remaining trapped atoms, collisions among
the outcoupled atoms, and the effect of the trapping potential.  We assume that
enough time has passed such that the outcoupled atom cloud from the left
condensate completely overlaps the right condensate. We hence neglect the time
evolution  of the outcoupled cloud while flying from the left to the right trap.
In our model,  this evolution leads to an additional phase factor which is
inconsequential to our findings. We also take the same functional form of the
trapping potential for the atoms in the left and right condensate such that
$\phi_b(\V{r}-\V{l})=\phi_c(\V{r}) \equiv \phi(\V{r})$ where
$\V{l}=\V{r}_R-\V{r}_L$ is the distance vector between the two potential minima.  
With these assumptions we find for the effective Hamiltonian
\begin{equation}\label{eqn:Heff}
H_{\rm eff}=H_A+H_{AL}+H_{\rm EM},
\end{equation}
where
\begin{equation}
H_{A} = \left [\delta\! \mu +\frac{\hbar \Omega_1^2}{\Delta}\right ] \hat{c}^\dagger \hat{c}
+\frac{\hbar \Omega_2^2}{\Delta} \hat{b}^\dagger_k \hat{b}_k +\frac{\hbar \Omega_1 \Omega_2}{\Delta} \left( \hat{c}^\dagger \hat{b}_k+ \hat{b}^\dagger_k \hat{c} \right )  \label{eq:A}
\end{equation}
describes the Rabi oscillations between the outcoupled atoms and the atoms in the
right condensate due to the Bragg beams. Here $\Omega_j$ are the Rabi frequencies
of the Bragg beams, $\Delta$ is the detuning from the excited state $\ket{e}$
which couples the two photon Raman transition between $\ket{b_k}\leftrightarrow
\ket{c}$ and the operator $\hat{b}_k$ annihilates an outcoupled atom in the
momentum shifted state $\ket{b_k}$ with wavefunction $\phi_k(\V{r})$.  
Hamiltonian (\ref{eqn:Heff}) is written in the reference frame of the Bragg beams
where we assume the two laser frequencies to be equal $\omega_1\approx
\omega_2=\omega_L$.  The term
\begin{eqnarray}
H_{\rm EM} &=&\hbar \sum_\lambda \Delta_\lambda \hat{a}_\lambda^\dagger \hat{a}_\lambda
\end{eqnarray}
takes into account the electromagnetic (EM) vacuum  energy, where we have used the standard plane wave
decomposition for the EM-field modes. Specifically, the positive frequency component of the vacuum electric field amplitude reads
\begin{equation}
\label{eqn:dEdef}
\delta\hat{\V{E}}^+(\V{r},t) = \sum_\lambda
\sqrt{\frac{\hbar \omega_\lambda}{2 \varepsilon_0\mathcal V}}
\pol_\lambda \hat{a}_\lambda(t) \rme^{\rmi \V{k}_\lambda\cdot\V{r}} \, .
\end{equation}
Here $\lambda$ labels a mode of
the EM-field at wavevector ${\bf k}_\lambda$, polarization $\pol_\lambda
\perp {\bf k}_{\lambda}$, and frequency $\omega_\lambda=c|{\bf k}_\lambda|$. The velocity of light is denoted
by $c$, the quantization volume by $\mathcal V$,  the vacuum permittivity is $\varepsilon_0$, and $\Delta_\lambda=\omega_\lambda-\omega_L$. The
operator $\hat{a}_{\lambda}$  annihilates a photon in mode $\lambda$.
The total electric field is the sum of the coherent Bragg laser fields
${\bf E}^+_{{\rm in},j}(\V{r})$ and the vacuum fields $\delta
\hat{\V{E}}^+(\V{r})$. The coherent part is responsible for the driving terms
in \eqref{eq:A}, while $\delta \hat{\V{E}}^+(\V{r})$ provides the coupling
of the vacuum modes with the atomic dipoles. We consider off-resonant
scattering where the scattering rates for sufficiently large condensates are
proportional to the amplitudes of the macroscopically occupied modes due to
Bose-enhancement, and we neglect scattering to other motional states of
the atoms. The coupling between the vacuum modes and the atoms is then given
by
\begin{eqnarray}
H_{AL} &=& \hbar \sum_\lambda \left (\hat{a}_\lambda^\dagger \hat{B}_\lambda+\hat{B}_\lambda^\dagger \hat{a}_\lambda \right ) \, ,
\end{eqnarray}
where we have introduced the operator
\begin{equation}\label{eqn:Bdef}
\hat{B}_\lambda=\Big((A_{1}^\lambda)^*  \hat{\sigma}_1+(A_{2}^\lambda )^*
\hat{\sigma}_2   \Big),
\end{equation}
with
\begin{equation}\label{eqn:defsig}
 \hat{\sigma}_1 = \hat{c}^\dagger \hat{e} , \qquad
\hat{\sigma}_2 = \hat{b}_k^\dagger \hat{e} \,,
\end{equation}
and, after adiabatic elimination, the excited state annihilation operator can be
written
\begin{equation}
\hat{e} = \left( \frac{\Omega_1}{\Delta}\hat{c}+\frac{\Omega_2}{\Delta}\hat{b}_k
\right ).
\end{equation}
We have defined
\begin{eqnarray}
A_{j}^\lambda &=& \sqrt{\frac{\hbar\omega_\lambda}{2\varepsilon_0 \mathcal V}}
\left(\V{d}^-_j\cdot\pol_\lambda\right) \int \rmd \V{r} |\phi(\V{r})|^2  \rme^{-\rmi
(\V{k}_j-\V{k}_\lambda) \cdot \V{r}} \, ,
\end{eqnarray}
where the factor outside the integral is the coupling strength between the atomic dipoles and the electromagnetic
field mode $\lambda$ \cite{Cohen}. Here
the matrix elements of the dipole moment operator $\hat{d}$ for the transition are
denoted  by $\V{d}^-_1 = \bra{c}\hat{d}\ket{e},\V{d}^-_2 = \bra{b_k}\hat{d}\ket{e}$.

\section{Continuous homodyne measurement} \label{contmeas}

We consider the condensate and the outcoupled atomic cloud together with the
driving fields as an open quantum system and eliminate the vacuum EM
field modes. The aim of our treatment is to compute the evolution of the reduced
system under continuous measurement of the light intensity of the Bragg beams.
The intensity of the beam $j$ is given by
\begin{equation} I_j=2c\varepsilon_0
\expct{\hat{\V{E}}_j^-(\V{r},t)\hat{\V{E}}_j^+(\V{r},t)} \,.
\end{equation}
Here the total electric field amplitude of each Bragg beam is given by
the sum of the coherent driving laser field and the field $\delta
\hat{\V{E}}_j^+(\V{r},t)$ due to scattering in the
direction of the beam $j$
\begin{equation}
\hat{\V{E}}_j^+(\V{r},t) = {\bf E}^+_{{\rm in},j}(\V{r})+\delta
\hat{\V{E}}_j^+(\V{r},t)\,.\label{totalfield}
\end{equation}
Assuming that the amplitude of the scattered field is small compared to the
applied laser field, the measured intensity is approximately
\begin{equation}\label{eq:intensityfluctuations}
I_j \simeq 2c\varepsilon_0\left(\expct{{\bf E}^-_{{\rm in},j}
{\bf E}^+_{{\rm in},j}}+
\expct{{\bf E}^-_{{\rm in},j}\delta \hat{\V{E}}_j^+}+
\expct{\delta \hat{\V{E}}_j^-{\bf E}^+_{{\rm in},j}}\right),
\end{equation}
where the last two terms give rise to fluctuations in the intensity incident on
the detector $j$.

We may now solve the intensity fluctuations by calculating the scattered field amplitude from the effective system Hamiltonian.
From the Heisenberg equation of motion for $\hat{a}_\lambda$ one finds
\begin{equation} \label{eqn:aHsbrg}
\fl \qquad \hat{a}_\lambda(t) =
\hat{a}_\lambda(0)\rme^{-\rmi \Delta_\lambda t}   -\rmi \int _0^t \rmd t'
\rme ^{-\rmi \Delta_\lambda(t-t')} \left( (A_{1}^\lambda)^* \hat{\sigma}_1(t')+
(A_{2}^\lambda)^* \hat{\sigma}_2(t')  \right )   \, ,
\end{equation}
with $\langle \hat{a}_\lambda(0)\rangle=0$. Inserting
(\ref{eqn:aHsbrg}) into (\ref{eqn:dEdef}) and defining
$\V{q}_j=k_L\V{n}-\V{k}_j$ as the change of the wavevector of light upon
scattering with $k_L=\omega_L/c$ \cite{Javanainen1995}, we then obtain two
contributions from the vacuum field, one for each beam
\begin{equation} \label{eqn:dE}
\delta \hat{\V{E}}_j^+(\V{r},t) \simeq  \frac{k_L^2 \rme^{\rmi k_L r_D} }{4
\pi\varepsilon_0 r_D} \V{n}\times(\V{n}\times \V{d}^-_j)
\hat{\sigma}_j(t)  \int \rmd \V{r}'
|\phi(\V{r}')|^2 \rme^{\rmi \V{q}_j \cdot \V{r}'} \, .
\end{equation}
The spatial integral over the wavefunction $\phi(\V{r}')$ enforces an
approximate momentum conservation, so that the photons are dominantly
scattered into a cone centred at $\V{q}_j=0$ in the direction of the laser
beam $j$. In deriving (\ref{eqn:dE}) we made the expansion
$|\V{r}-\V{r}'|=r_D-\V{n}\cdot \V{r}'$, with $\V{n}$ being the unit vector
that points from the scattering region to the detector, $r_D$ is the distance
between the detector and a representative point at the origin of the
scattering region.  Due to the normalization
of the wavefunction we finally find for the scattered electric field in the two
outgoing beams
\begin{eqnarray}\label{eqn:dEbeam} \delta
\hat{\V{E}}_j^+(\V{r},t) &=& \frac{k_L^2 \rme^{\rmi k_L r_D} }{4 \pi
\varepsilon_0  r_D} \V{n}\times(\V{n}\times \V{d}^-_j)
\hat{\sigma}_j(t)  \, .
\end{eqnarray}
The atomic operator associated with the
spontaneous emission of a photon into beam $j$ in (\ref{eqn:dEbeam}) is given by $\hat{\sigma}_j$. The
master equation which describes the evolution of the reduced density matrix
after elimination of the vacuum field modes then reads \cite{Carmichael_book}
\begin{equation}\label{eqn:mastereq}
\dot{\hat{\rho}}(t) = \frac{\rmi}{\hbar} [\hat{\rho},\hat{H}_A]-\sum_{j=1,2}
\frac{\gamma_j}{2} \left ( \hat{\sigma}_j^\dagger \hat{\sigma}_j
\hat{\rho}+\hat{\rho} \hat{\sigma}_j^\dagger \hat{\sigma}_j -2 \hat{\sigma}_j
\hat{\rho} \hat{\sigma}_j^\dagger \right )  \, .
\end{equation}
Here $\gamma_j$ is the rate of spontaneously
scattered photons, and is related to the total spontaneously scattered light
intensity $\delta I =2c\varepsilon_0 \expct{\delta \hat{\V{E}}_j^-(\V{r},t)
\delta \hat{\V{E}}_j^+(\V{r},t)}$ via \cite{MacSupJanne}
\begin{equation}
\label{eqn:Ideltatogam} \frac{1}{\hbar k_L c} \int \rmd \Omega
r_D^2 \delta I_j = \gamma_j \expct{\hat{\sigma}_j^\dagger \hat{\sigma}_j} \, ,
\end{equation}
where the angular integral is over the scattering cone of beam
$j$. The operators which are
associated with the light field amplitude of the beam $j$ read
\begin{equation}
\hat{C}_j=\sqrt{\gamma_j}(\alpha_j+\hat{\sigma}_j),
\end{equation}
where $\alpha_j$ is proportional to the amplitude of the coherent laser beam
with wavevector $\V{k}_j$.  The intensity is then proportional to
$\expct{\hat{C}_j^\dagger\hat{C}_j}$. The leading contribution comes from the
coherent intensity $\propto \gamma_j |\alpha_j|^2$ [corresponding to the first
term in \eqref{eq:intensityfluctuations}] and the intensity fluctuations
are dominated by the terms $\gamma_j (\alpha_j \langle \hat{\sigma}_j\rangle
+{\rm C.c.})$ [corresponding to the second and the third term in
\eqref{eq:intensityfluctuations}]. Extending the treatment of
\cite{Wiseman_homodyne} for our setup one finds that the evolution of the
system under the continuous monitoring of light intensity can be
described by the stochastic differential equation
\begin{equation}\fl \label{eqn:Stoch} \quad \ket{\psi(t+\rmd
t)}=\left(1-\frac{\rmi}{\hbar}
\hat{H}_A \rmd t+\sum_j \left[\frac{\gamma_j}{2} \hat{\sigma}_j^\dagger
\hat{\sigma}_j \rmd t + 2\gamma_j X_j \rmd t +\hat{\sigma}_j \sqrt{\gamma_j}
\rmd W_j \right ]  \right)
\ket{\psi(t)} ,
\end{equation}
where
\begin{eqnarray} X_j &=& \frac{1}{2} \left
(\hat{\sigma}_j+\hat{\sigma}_j^\dagger \right ).
\end{eqnarray}
Here $\rmd W_j$ is a Wiener increment with zero mean $\langle \rmd
W_j\rangle=0$ and $\expct{(\rmd W_j)^2}=\rmd t$, which appears as a result of
the continuous measurement process.   Keeping
terms to lowest order in the fluctuations one finds an expression for the
photocurrent in essence equivalent to (\ref{eq:intensityfluctuations})
\begin{eqnarray}
\label{eqn:Imeas} i^{\mathrm{phot}}_j(t) &=& \gamma_j
\alpha_j^2 +\alpha_j \left( 2 \gamma_j \expct{X_j} +\sqrt{\gamma_j} \, \xi_j(t)
\right ) \,.
\end{eqnarray}
Here
\begin{equation} \xi_j(t)=\frac{\rmd W_j}{\rmd t},
\end{equation}
represents Gaussian white noise \cite{Wiseman_homodyne} and arises from the open
nature of our quantum system.

\section{Numerical Results} \label{numres}

In order to study the effect of the homodyne photo-current measurements on the
system, we numerically integrate (\ref{eqn:Stoch}), using the Milstein
algorithm~\cite{GardinerHandbook}.  As an initial state in the numerical
simulations we take a pure number state in each condensate, with no
well-defined phase between them, and the incident Bragg laser beams are taken to
be classical coherent states.  The relative phase between the condensates
as a function of time may then be calculated as $\Phi(t) = \mbox{arg}(\langle
\hat{c}^\dagger\hat{b}\rangle)$. We define a measure for the strength of
the phase coherence between the condensates by the absolute value of the
normalized phase coherence
\begin{equation}
g(t) = \frac{|\langle \hat{c}^\dagger\hat{b}\rangle|}
{\sqrt{\langle \hat{c}^\dagger\hat{c}\rangle
\langle \hat{b}^\dagger\hat{b}\rangle}}.
\end{equation}
A value of $g(t)$ close to one indicates a high degree of relative
phase coherence, while condensates with no relative phase information have
$g(t)\simeq 0$.
\begin{figure}[htp] \centering
\includegraphics[width=0.8\textwidth]{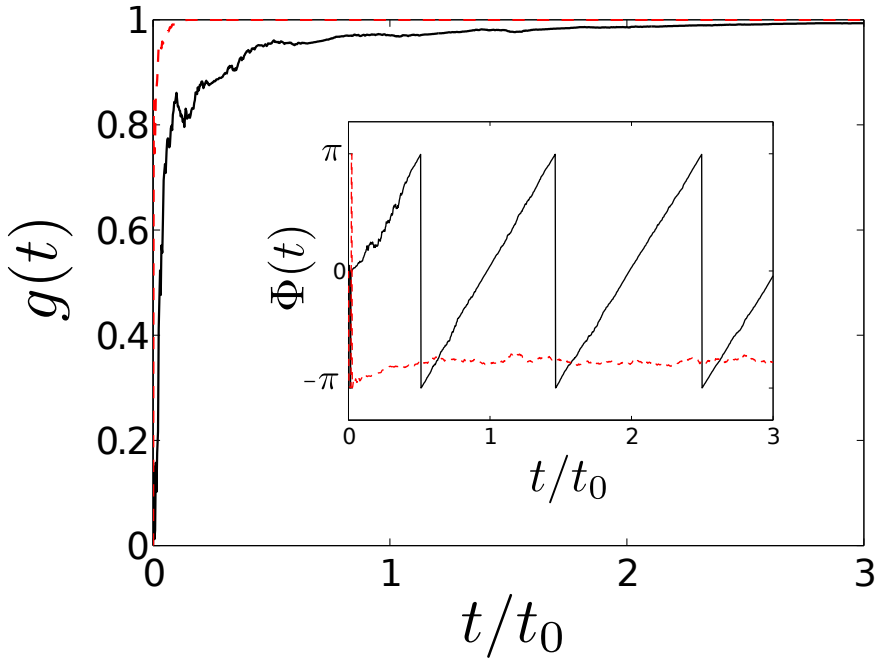}
\caption{ \label{fig:coherence} (colour online) Coherence $g(t)$ as a function
of time $t$ in units of $t_0=2\pi \hbar/\delta\! \mu$. The simulation is done
for a total of $N=100$ atoms which are initially distributed equally between
the states $\ket{b_k}$ and $\ket{c}$. Parameters were chosen to be
$\gamma_1=\gamma_2 = 10^5/t_0$ and $\sqrt{\Omega_1\Omega_2} = 10^3/t_0$. The black
solid line corresponds to $\Delta=100 \gamma_1$ and the red dashed line to
$\Delta=10 \gamma_1$. In the inset we show the evolution of the phase $\Phi(t)$
as a function of time $t$ in units of $t_0$.
}
\end{figure}

\begin{figure}[htp] \centering
\includegraphics[width=0.8\textwidth]{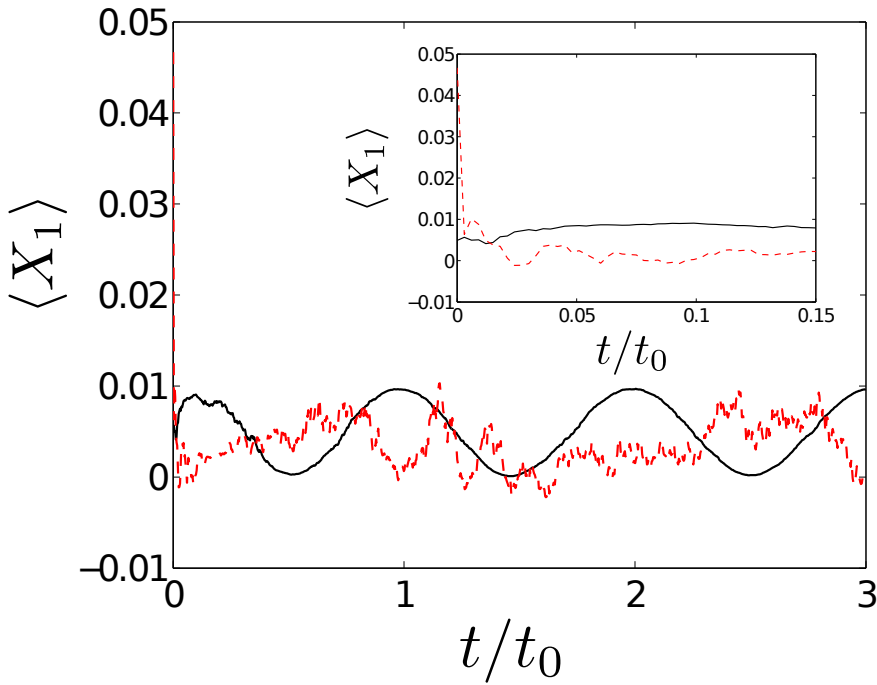}
\caption{ \label{fig:X1} (colour online) Mean value of the quadrature
$\expct{X_1}$ as a function of time $t$.  The black solid line corresponds to
$\Delta=100 \gamma_1$ and the red dashed line to $\Delta=10 \gamma_1$. Other
parameters are as in figure~(\ref{fig:X1}).  The inset shows an expanded view of
the initial behaviour.
}
\end{figure}

In figure~\ref{fig:coherence} we plot the time evolution of the coherence and the
relative phase for two different values of the detuning $\Delta$. No
well-defined relative phase exists at early times, the coherence starts at zero and
$\Phi(t)$ shows large random fluctuations with time. As
the continuous measurement proceeds the coherence builds rapidly, leading to a
well-defined relative phase with a stable value.
Once established, we then see two different regimes of behaviour at longer times.
For large values of the detuning $\Delta=100\gamma_1$ we see a running phase
behaviour: once well established with a value which is random for each
individual run, the phase grows linearly in time with a rate
proportional to the difference in energies between the condensates
$\Phi(t)\sim\delta\! \mu \,t/\hbar$.  From (\ref{eqn:Imeas}) we note that the
measured photocurrent from the two Bragg beams is essentially proportional to
the quadrature $\expct{X_j}$ after subtracting the background current, and the
corresponding time evolution of $\expct{X_j}$ is shown in figure~\ref{fig:X1}.  In the
running phase regime, the quadrature exhibits well defined oscillations with
frequency $\omega_{\rm osc}=\delta \mu/\hbar$, and this corresponds to the
experimental measurements obtained by Saba {\it et al.\,} \cite{Saba}.  Such
oscillations thus give an interferometric measurement of the relative phase
evolution, sensitive to any accumulated phase shift due to an energy offset between
the distant condensates.  An interferometer of this type could be used, for example, to detect
a weak force applied to one of the condensates.

Choosing a smaller value for the detuning $\Delta=10\gamma_1$ we find a very
different long time behaviour.   Once again the phase fluctuates as coherence is
established, although this occurs on a much faster timescale.  This can be
understood from the fact that the phase is established as a result of the intensity
fluctuations in the laser beams, which are enhanced by decreasing the detuning
$\Delta$.  Unlike the running phase case, once firmly established the phase now
locks to an almost constant value near to $\pi$.  This trapped phase state has
the two condensates almost entirely out of phase, leading to destructive
interference in the oscillations between the states $\ket{b_k}$ and $\ket{c}$  and
resulting in a state analogous to a dark-state. The corresponding quadrature
$\expct{X_1}$ therefore exhibits merely random fluctuations which would not be
suited to an interferometric type experiment.  Note that while one would expect the
amplitude of these random fluctuations to be suppressed compared to the coherent
oscillations of the running phase regime, and this is indeed the case, the two
different values of detunings used here do not allow such a direct comparison in
figure~\ref{fig:X1}.

The two different regimes of behaviour resemble the AC-Josephson and
self-trapping behaviours seen in double-well
condensates~\cite{Raghavan1999,Albiez2005,Gati2007}, although we emphasize that
here the coupling occurs due to the non-local measurement process induced by the
Bragg beams.  The trapped phase behaviour is more akin to a dark state, however,
due to the lack of any nonlinearity in the Hamiltonian which is required for
macroscopic self-trapping.  The different regimes may be understood if we assume that a well
defined phase and population can be associated with each condensate, and neglect
any processes other than those included in $H_A$, leaving a two-mode model
similar to that considered in \cite{Raghavan1999}.  The trapped phase regime
then occurs with a stable relative phase difference of $\pi$ when
\begin{equation}
\delta\!\mu = 2\hbar \frac{\Omega_1\Omega_2}{\Delta}\frac{|z|}{\sqrt{1-z^2}},
\label{eq:darkstatecondition}
\end{equation}
where $z = (N_{k}-N_c)/(N_{k}+N_c)$ is the relative population difference,
where $N_{k(c)}$ is the population in state $\ket{b_k}(\ket{c})$.  The
trapped phase regime therefore requires either $\delta\!\mu/\hbar \sim
{\Omega_1\Omega_2/\Delta}$, or a large population imbalance.  This is in agreement with
our results, where the larger value of detuning has
$\delta\!\mu/\hbar\gg{\Omega_1\Omega_2/\Delta}$, and the initial population balance is
not extreme.  The trapped phase condition is then not satisfied and we observe a running
phase behaviour akin to the AC-Josephson effect.  Note, however, that the dissipation is
vital for establishing the relative phase in the first place, and has the effect of
shifting the relative phase in the trapped regime away from $\pi$ in
figure~\ref{fig:coherence}. The model (\ref{eq:darkstatecondition}) does not specify
the role of spontaneous scattering which determines the rate at which the system is driven towards
the trapped phase state.

In the experiment of \cite{Saba}, the system typically contained on the order of
$10^6$ atoms, although numbers outcoupled would be only a small fraction of
this. Numerical simulations in our basis for such large numbers are
prohibitively slow, so here we have typically used a total atom number of $100$.
Our results show no significant dependence on atom number, however, and we
expect our results to give a good qualitative comparison with the physics
exhibited in the experiment.  We have so far discussed our results in
dimensionless units; in order to give a specific example we may take
$\gamma_{1,2}$  to be of the order of $2\pi\times10\mbox{MHz}$, for instance.
Our results then correspond to the values
$\delta\!\mu/\hbar\sim2\pi\times630\mbox{Hz}$, and
$\sqrt{\Omega_{1}\Omega_2}\sim2\pi \times0.45\mbox{MHz}$.

Parameters used in the experiment of \cite{Saba} were  $\delta\!\mu/\hbar \sim
2\pi\times1\mbox{kHz}$, $\Delta = 2\pi \times 1\mbox{GHz}$,
$\sqrt{\Omega_{1}\Omega_2} \sim 2\pi\times0.45\mbox{MHz}$. This yields a ratio
$\eta \equiv 2\hbar\Omega_1\Omega_2/(\Delta \delta\!\mu)\sim 0.4$.  In order for
the trapped phase regime to be observed the population imbalance would then be
required to satisfy $z \approx 0.92$.  In the experiment, the actual population in
the momentum shifted state ($\ket{b_k}$) that overlapped the second condensate
($\ket{c}$) was of the order of $2\times 10^4$ atoms during the coupling.  The
corresponding population imbalance was in excess of $0.96$, and hence did not
satisfy condition (\ref{eq:darkstatecondition}) for the trapped phase behaviour.
The observed running phase behaviour in the experiment is therefore consistent
with our model.

\section{Concluding remarks} \label{Sum}

Bragg spectroscopy was used in \cite{Saba} to measure the relative phase
between two initially uncorrelated BECs.  By studying a simplified model containing
the essential ingredients of the experiment, we have demonstrated how the homodyne
measurement process builds up a coherent relative phase between the two
condensates.  This quantum measurement-induced backaction entangles the two
macroscopic many-body states even though the measurement location can be far removed
from the region of interactions.

Following the establishment of a coherent phase, we have identified two distinct
behaviours under continual subsequent measurement.  With a larger atom-laser detuning
$\Delta$, or a large initial energy imbalance $\delta\!\mu$, we reproduce the
experimental findings of \cite{Saba}, with the measured photon flux exhibiting
oscillations at a frequency corresponding to the energy offset of the
separated condensates.  In this case, once the coherence and a random-valued phase is established it
evolves linearly in time $\Phi(t)\sim\delta\!\mu t/\hbar$. Measurable
oscillations in the laser beam intensity mean this state has applications to
quantum-enhanced interferometry, and the measurement backaction could potentially
be used further to implement feedback mechanisms~\cite{Feedback}.
By choosing a smaller atom-laser detuning, we
found instead that the system stabilized to a trapped phase state with the
condensate relative phase fixed at almost $\pi$, while the scattered light
intensity showed only random fluctuations.  A semiclassical model can
qualitatively describe the difference between these two regimes, with the
trapped phase behaviour occurring when (\ref{eq:darkstatecondition}) was
satisfied.

When atoms from two initially uncorrelated condensates overlap, we have shown that
Bragg coupling and continuous homodyne measurement can rapidly establish a
well-defined relative phase.  A closely related experiment~\cite{Ginsburg2007} has
been performed using ultra-slow light pulses, in which optical information was
coherently transported between two spatially separated condensates by a travelling
matter wave.  An ultra-slow light pulse was stopped in the first condensate,
creating a dark-state superposition between two atomic internal states.  Upon
stopping the pulse one of the internal states received a momentum kick, and
outcoupled atoms in this state passed through a second distant condensate.  By
illuminating the second condensate with a coupling laser it was possible to revive
the initial light pulse, even when the BECs were independently produced.  In the case where the condensates had been prepared
separately, a rapid establishment of a coherent phase in a manner similar to that
described in this paper explains the recovery of the light pulse.

\ack The authors acknowledge G. Morigi and S. Pugnetti for stimulating discussions
and helpful comments. This work was supported by the European Commission (EMALI) and the Leverhulme Trust.  The research leading to these results has
received funding from the European Union FP7/ 2007-2013 under grant agreement N.
234970 - NANOCTM.

\end{document}